\documentclass[aip,preprint]{revtex4-1}
\usepackage{graphicx}% Include figure files
\usepackage{dcolumn}% Align table columns on decimal point
\usepackage{bm}% bold math
\usepackage[mathlines]{lineno}% Enable numbering of text and display math
\usepackage[utf8]{inputenc}
\usepackage[T1]{fontenc}
\usepackage{mathptmx}
\usepackage{etoolbox}
\usepackage{braket}
\usepackage{multirow}
\usepackage{trimclip}
\usepackage{amsmath}
\usepackage{accents}
\usepackage{placeins}

\makeatletter
\def\@email#1#2{%
 \endgroup
 \patchcmd{\titleblock@produce}
  {\frontmatter@RRAPformat}
  {\frontmatter@RRAPformat{\produce@RRAP{*#1\href{mailto:#2}{#2}}}\frontmatter@RRAPformat}
  {}{}
}%
\makeatother
\newcommand{\fref}[1]{Figure~\ref{#1}}

\newcommand{\tref}[1]{Table~\ref{#1}}
\DeclareUnicodeCharacter{2032}{$'$} % ′ PRIME → TeX math prime

\begin{document}

%\preprint{AIP/123-QED}

\title[]{Kinetics of Barrier Crossing Events from  Temperature Accelerated Sliced Sampling Simulations}
% Force line breaks with \\
\author{Sameer Saurav}
\altaffiliation[Also at ]{Department of Chemistry, Anugrah Narayan Smarak College Nabinagar, 824301, India.}%Lines break automatically or can be forced with \\
\affiliation{ 
Department of Chemistry, Indian Institute of Technology Kanpur (IITK), Kanpur, 208016, India.%\\This line break forced with \textbackslash\textbackslash
} 

\author{Debjit Das}
\affiliation{Department of Chemistry, Indian Institute of Technology Bombay (IITB), Mumbai, 400076, India.}

\author{Ramsha Javed}
\author{Nisanth N. Nair}

% \homepage{https://home.iitk.ac.in/~nnair/}
\affiliation{ 
Department of Chemistry, Indian Institute of Technology Kanpur (IITK), Kanpur, 208016, India.%\\This line break forced with \textbackslash\textbackslash
}%
  \email{nnair@iitk.ac.in}
\date{\today}% It is always \today, today,

\begin{abstract}
Temperature-accelerated sliced sampling (TASS) is a well-established enhanced sampling method that facilitates exhaustive exploration of high-dimensional collective variable (CV) space through directed sampling employing a combination of umbrella restraining biases, metadynamics biases, and temperature acceleration of CVs. 
In this work, we broaden the applicability of TASS by introducing a protocol for computing rate constants of barrier crossing events. 
The challenge addressed here is to recover kinetics from free energy data computed from different slices of the TASS simulation.
The proposed protocol utilizes artificial neural networks based representation of high-dimensional free energy landscapes, and Infrequent Metadynamics.
We demonstrate the accuracy of the approach by obtaining rate constants for the conformational change of alanine dipeptide {\em{in vacuo}}, the unbinding of benzamidine from trypsin, and the unbinding of aspirin from $\beta$-cyclodextrin.
\end{abstract}
% We demonstrate that our method recovers the correct transition rate.
\maketitle

\section{\label{sec:level1}Introduction}
Enhanced sampling molecular dynamics (MD) techniques are widely used to accelerate barrier crossing events in simulations and  compute free energy surfaces for these processes.\cite{tuckerman2023statistical,frenkel2023understanding,peters2017reaction,vanden2009some,christ2010basic,bonella2012theory,valsson2016enhancing}
Linking these simulations directly to experiments requires calculating reaction rates using the free energy data.
Many computational methods have been developed to predict reliable reaction kinetics, going beyond the direct application of Eyring’s equation, which connects free energy barriers to rate constants.\cite{ribeiro2018kinetics,sohraby2023advances,bernetti2017protein,decherchi2020thermodynamics,bernetti2019kinetics}
Methods like milestoning, \cite{faradjian2004computing,bello2015exact,elber2020milestoning} adaptive multiple splitting, \cite{cerou2007adaptive,cerou2011multiple,teo2016adaptive}transition path sampling,  \cite{dellago1998transition,vanden2006towards,dellago2006transition,escobedo2009transition, juraszek2013efficient} and weighted ensemble \cite{huber1996weighted,bhatt2010steady,zhang2010weighted,zuckerman2017weighted,dickson2016ligand,dickson2017multiple,nunes2018escape} approaches use statistics from an ensemble of trajectories.
Other approaches based like infrequent metadynamics (IMetaD),\cite{tiwary2013metadynamics,salvalaglio2014assessing} $\tau$-RAMD,\cite{kokh2018estimation} Ligand Gaussian accelerated MD, \cite{miao2020ligand} and dissipation-corrected targeted MD\cite{wolf2018targeted} are built for obtaining kinetics from biased simulations. 

In this work, we have developed an approach for calculating rates constants using temperature accelerated sliced sampling (TASS). \cite{awasthi2017exploring,awasthi2019wiley,paul2019phase,bajpai2023interoperable,javed2024buckets}
TASS is an enhanced sampling technique that ensures exhaustive exploration through a controlled-directional sampling and temperature acceleration.
The method is built using the dynamic Adaptive Free Energy Dynamics (d-AFED) framework \cite{maragliano2006temperature,abrams2008efficient} which enables the use of a large number of CVs. 
The high temperature auxiliary variables coupled to the CVs accelerate the diffusion of the system in the CV space.
The directionality of the sampled rare event is achieved through an umbrella bias along a CV, enhancing sampling efficiency without the system becoming trapped in a high-entropy. 
In addition, metadynamics \cite{laio2002escaping,barducci2008well,iannuzzi2003efficient} or parallel-bias metadynamics \cite{pfaendtner2015efficient} type biases can be applied to a selected set of CVs. 
This method has previously been employed to determine the free energy of various chemical and biological processes, including protein folding, \cite{bajpai2023interoperable,javed2024buckets} enzymatic reactions, \cite{awasthi2018mechanism,vithani2018mechanism,soniya2019transimination,thakkur2022inhibition} protein-ligand unbinding, \cite{tripathi2023temperature} membrane permeation, \cite{acharya2021improved,acharya2023antibiotic,acharya2024molecular,acharya2025improved} and catalytic reactions in zeolites. \cite{verma2022proton} 
Given that the enhanced sampling is conducted independently to generate biased distributions with different umbrella biases, applying existing strategies to compute reaction rates becomes challenging. 
This work put forward an approach to overcome this using the concept of IMetaD.

The manuscript is organized as follows. 
We first discuss the protocol for calculating kinetics under the Methods section. 
Subsequently, we demonstrate the performance of the method by computing the rates for some of the well studied systems: Conformational change of alanine dipeptide  \textit{in vacuo}, the unbinding of benzamidine from trypsin, and the unbinding of aspirin from $\beta$-cyclodextrin ($\beta$-CD).

\section{\label{sec:level2}Method}
\label{sec:method}
\subsection{Temperature Accelerated Sliced Sampling}

The TASS Lagrangian is given by

\begin{equation}
\label{TASS}
    \begin{split}
    {\mathcal{L}}_{h} (\mathbf{R}, \dot{\mathbf{R}}, \mathbf{s}, \dot{\mathbf{s}}) & =  {\mathcal{L}}^{0} (\mathbf{R}, \dot{\mathbf{R}}) + \sum_{{\alpha}=1}^{n}  \left[ {\frac{1}{2} \mu_{\alpha} {\dot{s}}_{\alpha}^2} - \frac{{\kappa}_{\alpha}}{2} (q_{\alpha}(\mathbf{R}) - s_{\alpha})^2 \right] \\
& \quad   - {W}_{h}^{\mathrm b} ({s_1}) - V^{\mathrm b}({\mathbf s}^{\mathrm m},t), 
 \enspace h=1,\ldots,M.
    \end{split}
\end{equation}
where ${\mathcal{L}}^{0} (\mathbf{R}, \dot{\mathbf{R}})$ is the Lagrangian for the original system. 
Here, $\mathbf{R}$ and $\dot{\mathbf{R}}$ are the set of atomic coordinates and velocities, respectively, while $\mathbf{s}$ and $\dot{\mathbf{s}}$ represent the auxiliary variables and their velocities. 
In the above, $n$ number of CVs, $\left \{ q_\alpha(\mathbf R) \right \}$, are defined.
The auxiliary variables are assigned 
a mass, $\mu_{\alpha}$, and  ${\kappa}_{\alpha}$ defines the spring constant of the coupling potential between the physical and the auxiliary variables. 
The auxiliary variables are maintained at a higher temperature to ensure enhanced conformational sampling, while the physical subsystem is kept at 300~K. 
The parameters $\mu_{\alpha}$ and 
${\kappa}_{\alpha}$
are chosen to keep the physical and auxiliary spaces adiabatically decoupled. 
%
%An advantage of this method is that the simulation time does not increase exponentially with the number of CVs. 

The bias  ${W}_{h}^{\mathrm b} ({s_1})$ is a harmonic bias,  defined as
\begin{equation}
     {W}_{h}^{\mathrm b} ({s_1}) = \frac{1}{2} k_h \left[ s_1 (\mathbf{R}) - {\xi}_h \right]^2, \enspace h=1,\ldots,M.
\end{equation}
This bias is usually applied along one of the $n$ auxiliary variables, say $s_1$, and are centered at $M$ different values, ${\xi}_h,\enspace h=1,..., M $. 
The term $V^{\mathrm b}({\mathbf s}^{\mathrm m},t)$ (Equation \ref{vbias:mtd}) is a well-tempered metadynamics (WTMetaD) bias \cite{barducci2008well,dama2014well} that can be applied along a subset of auxiliary variables ${\mathbf s}^{\mathrm m}$:

\begin{equation}
\label{vbias:mtd}
    V^{\mathrm b}(\mathbf s^{\mathrm m},t) = \sum_{\tau<t} w_\tau\,\exp{\left[-\frac{{\left\|\mathbf s^{\mathrm m} -  {\mathbf s}_\tau^{\mathrm m} \right\|}^2}{2(\delta s)^2}\right]}
\end{equation}
with
\begin{equation}
 w_\tau = w_0\exp\left[{-\frac{V^{\mathrm b}( {\mathbf s}_\tau^{\mathrm m}, \tau)}{k_{\mathrm B}\Delta T}}\right], 
 \end{equation}
and $ {\mathbf s}_\tau \equiv \mathbf s(\tau)$ as in WTMetaD.\cite{barducci2008well,dama2014well} Here, $\tau$ represents a quantized time, and the Gaussian potentials are updated incrementally. The height of the gaussian deposited at time $\tau$ is $w_\tau$, and the width of the gaussian is $\delta s$, and the parameter that modulates the change of the gaussian height is $\Delta T$ (in Kelvin). The bias factor $\gamma$ is defined as 
\[ \gamma = \frac{(T + 
\Delta T)}{T} \enspace .
\]

Recently, a variant of TASS was proposed in which bucket potentials replace umbrella biases \cite{javed2024buckets}. 
This approach significantly reduces the number of umbrella biases needed in TASS simulations.
The Lagrangian in this case has all the terms the same as in equation \ref{TASS}, except the bias potential ${W}_{h}^{\mathrm b} ({s_1})$, which is now defined as,
\begin{equation}    
W_h^{\mathrm b}(s_1) = 
\begin{cases}
k_h \bigl(s_1 - \xi_h^{\mathrm L}\bigr)^4, & \text{if } s_1 < \xi_h^{\mathrm L} \\[6pt]
k_h \bigl(s_1 - \xi_h^{\mathrm U}\bigr)^4, & \text{if } s_1 > \xi_h^{\mathrm U} \\[6pt]
0, & \text{otherwise,}
\end{cases}
\end{equation}
where $k_h$ is the curvature of the wall potential, while $\xi_h^{\mathrm L}$ and $\xi_h^{\mathrm U}$ are the lower and upper limits for the CV $s_1$ in the $h^{\text{th}}$ window. 

\subsection{Artificial Neural Network (ANN) Representation  of the Free Energy Surfaces}
 
High-dimensional free energy surfaces obtained from TASS can be represented by an artificial neural network (ANN). \cite{cendagorta2020comparison,schneider2017stochastic} 
For an ANN with $K$ hidden layers and $M$ nodes, the free energy surface (FES) is represented as
\begin{equation}
    \begin{split}
     F(\textbf{s};w) =  H \bigg[ \sum_{j_k =1}^{m_k} h \bigg( ...h \bigg\{ \sum_{j_2 =1}^{m_2} h \bigg[ \sum_{j_1 =1}^{m_1} h \bigg( \sum_{\alpha =1}^{n}  s_{\alpha} w_{\alpha, j_1}^0 + w_{0, j_1}^0 \bigg) \\
    &  w_{j_1, j_2}^1 + w_{0, j_2}^1 \bigg]  w_{j_2, j_3}^2 + w_{0, j_3}^2 \bigg\} ...\bigg)  w_{j, k}^K +  w_{0}^K \bigg] .
    \end{split}
\end{equation}
where $w$ is a set of fitting parameters. The parameter $w_{ik}^{\nu}$ connects node $i$ of layer $\nu$ with node $k$ of layer  $\nu + 1$. 
The activation functions considered here are $H(x) = x$ and $h(x) = 1/(1+x^2)$.
If there are $N_{\rm g}$ free energy values $F^{(\lambda)}$ at CV values $s^{(\lambda)}$ are taken, the optimal set of parameters $w$ is obtained by minimizing the cost function
\begin{equation}
    E(w) = \frac{1}{2N_{\mathrm g}} \sum_{\lambda = 1}^{N_\mathrm g}\left( F(\mathbf{s}^{(\lambda)}; w) - F^{(\lambda)} \right)^2,
\end{equation}
%
%This cost function is minimized with respect to $w$ 
for the specified training set.
In this work, the ANN parameters were optimized using the adaptive moment estimation optimization (ADAM) algorithm \cite{kingma2014adam} and implemented using the Pytorch \cite{paszke2019pytorch} library of Python. 

\subsection{Infrequent Metadynamics (IMetaD)}
\label{sec:imetad}

Tiwary and Parrinello proposed the IMetaD method to predict kinetics within the framework of metadynamics simulations.\cite{tiwary2013metadynamics,salvalaglio2014assessing}
In this approach, the kinetics of a barrier-crossing process is obtained based on the first passage times measured for multiple independent metadynamics simulations with a very slow bias deposit rate.
The ratio of the first passage time in the unbiased  and the biased simulations defined as the acceleration factor $(\chi)$. 
While using the metadynamics bias, as in Eqn.~\ref{vbias:mtd},
\begin{equation}
    \chi = \langle e^{{\beta}V^{\rm b}(\mathbf{s},t)} \rangle,
\end{equation}
%where $\left < \cdots \right >$ denotes the time average carried out over the entire the simulation time. 
where $\left < \cdots \right >$ denotes the ensemble average.

Since a single simulation is insufficient for accurate rate estimation, multiple (typically 10 or more) \cite{salvalaglio2014assessing,tiwary2015kinetics,shekhar2022protein} simulations were performed with different initial conditions. 
Each simulation is terminated as soon as the product region is reached. 
The unbiased first passage time is then calculated by multiplying the simulation time $\chi$. 
As the transition is a rare event, the set of first passage times obtained from the independent simulations are expected to follow a Poisson distribution.\cite{resnick1992adventures,salvalaglio2014assessing} 
The corresponding {empirical} cumulative distribution function (ECDF) is estimated from the computed unbiased transition times and compared to the theoretical CDF (TCDF),
\begin{equation}
\label{eqn_acceleration}
    P_{n \geq 1} = 1- \enspace e^{-t/\tau}, 
\end{equation}
where $P_{n \geq 1}$ is the probability of observing at least one transition in time $t$, and 
$\tau$ is the transition time, which will be determined by fitting the ECDF with TCDF.
The reciprocal of $\tau$ gives the transition rate. 
To assess the reliability of the calculated rate, a two-sample Kolmogorov-Smirnov (KS) test \cite{massey1951kolmogorov,miller1956table} is performed between the TCDF and the ECDF. 

\subsection{Extending IMetaD to TASS Simulations}
\label{Steps_for_rate}
The ideas used in the IMetaD approach can be extended to TASS simulations for computing transition rates using the free energy surfaces obtained from TASS.
The fundamental idea will be to build a bias potential that nearly fills the reactant basin, knowing the information of the free energy surface from TASS.
Subsequently, an IMetaD simulation is  performed, starting the simulations with the bias constructed from TASS.
More detailed steps involved are listed below:
\begin{enumerate}
    \item  Compute $F(\mathbf s)$ from TASS simulations; see Refs.~\cite{awasthi2017exploring,pal2021mean} for the strategies to be followed here;
     \item Train an ANN to represent $F(\mathbf s)$
     \item Knowing the analytical form of $F(\mathbf s)$ as ANN, compute the bias $V^{\rm b}_0(\mathbf s)$ that nearly fills the reactant basin of $F(\mathbf s)$; In practice, we choose  $V^{\rm b}_0(\mathbf s)$ that is 90\% of the transition barrier;
     \item Start IMetaD, taking $V^{\rm b}_0(\mathbf s)$ as the initial bias. Run IMetaD simulation, i.e., well-tempered metadynamics (WTMetaD) with very slow bias deposition rate until a barrier crossing is seen in the simulation; The simulation time $\tau^{\text{b}} \equiv n_\tau \,  \delta t$ is recorded, where $n_\tau$ is the number of MD steps, and $\delta t$ is the MD time step;
     % Now calculate the acceleration factor ($\alpha$) (see Equation \ref{eqn_acceleration}).
     % \item Estimate the current bias potential $V(\mathbf{s_i})$ and compute $e^{{\beta}V(\mathbf{s_i})}$, at every step. $\mathbf{s_i}$ denotes the system’s position in the CV space at the $i^{\text{th}}$ step. At the first step, include $e^{{\beta}V_0(\mathbf{s_i})}$, where $V_0$ is the bias from step (3). 
     \item Compute the acceleration factor:
     \begin{eqnarray}
          \chi = \frac{1}{n_\tau} \sum\limits_{i}^{n_{\tau}} 
          \exp \left [ \beta V^{\mathrm b} \left (\mathbf{s}; t_i \right ) \right ] \, ;
     \end{eqnarray}
     \item Compute the unbiased transition time, $\tau^{\text{u}}$, using
     \begin{equation}
         \tau^{\text{u}} = \chi \, \tau^{\text{b}} \, ;
     \end{equation}
     \item Perform multiple simulations with the same initial structure but different velocities; 
     \item Compute the ECDF from the computed unbiased transition times and fit to TCDF to estimate $\tau$;
     %, where $\tau$ represents the mean first passage time (MFPT).
     \item Obtain the rate constant by taking the reciprocal of $\tau$:
     \begin{equation}
         k = \frac{1}{\tau} \, ;
     \end{equation}
     \item Perform a two-sample KS test to assess the reliability of the kinetics.
\end{enumerate}

\begin{figure}
\begin{center}
\includegraphics[width=14.0cm]{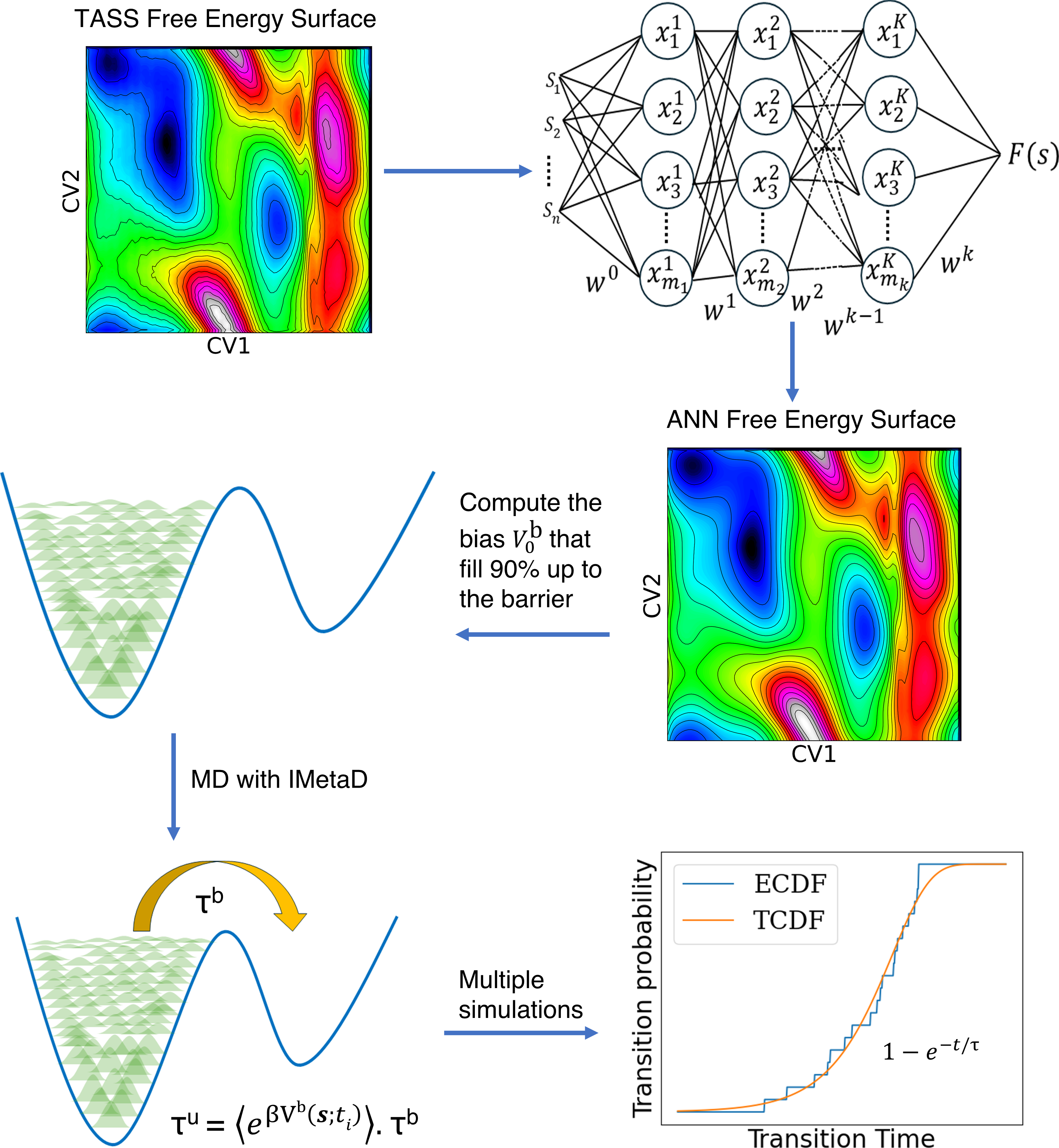}
\caption{Steps for calculating the barrier crossing rate employing IMetaD based on TASS data are outlined; See Section \ref{Steps_for_rate}. }
\label{fig:rate_method}
\end{center}
\end{figure}

For a schematic of the steps involved, see \fref{fig:rate_method}.
We obtain $V_0^{\rm b}$ in Step~3 as a linear combination of Gaussian functions.
To obtain the best linear combination, we perform Langevin dynamics of the variables $s_1, \cdots, s_n$ using $F(\mathbf s)$ as the potential and adding a well-tempered metadynamics bias  along $s_1, \cdots, s_n$. 
From the bias potential obtained in this simulation, it is straight forward to construct $V_0^{\rm b}$.
Representation of $F(\mathbf s)$ by ANN makes it easy to use that as the potential to perform Langevin dynamics.

\section{Computational Details}
\subsection{Alanine Dipeptide In \textit{Vacuo}}

Alanine dipeptide in \textit{vacuo} was modeled using the ff14SB AMBER force field.\cite{maier2015ff14sb}
MD simulations were carried out using the AMBER18 package \cite{case2018amber} patched with the PLUMED interface. \cite{bonomi2009plumed,tribello2014plumed,plumed2019promoting}
A time step of 1~fs was used to integrate the equations of motion. 
The Ramachandran angles $\phi$ and $\psi$ are taken as the CVs in TASS simulations; see SI~Section~S1.
An umbrella bias was applied along $\phi$ with $k_h = 239$~kcal~mol$^{-1}$~rad$^{-2}$ for all the  windows.
A WTMetaD bias was applied along the $\psi$ dihedral with $w_0$ = 0.57~kcal~mol$^{-1}$, $\delta s = 0.05$~rad and $\gamma = 10$.
The gaussian bias deposition stride was 0.5~ps. 
TASS parameters $\mu_\alpha$ and $\kappa_\alpha$ were 50~Da~{\AA}$^2$~rad$^{-2}$ and 1258~kcal~mol$^{-1}$~rad$^{-2}$, respectively. Both the physical ($T$=300~K) and extended ($\tilde T$=1000~K) systems were controlled using massive thermostatting with a Langevin thermostat, taking a friction coefficient of 0.1~$\mathrm {fs}^{-1}$. 
The mean-force-based approach as in Ref.~\cite{pal2021mean} was used for free energy reconstruction from the TASS simulation. 

%\subsubsection{IMetD setup}

We computed the TASS free energy surface along $(\phi,\psi)$ on a $ 100 \times 100$ grid. We trained an ANN with 80 \% of the free energy grid data, while another 20 \% of the data was used for testing. 
The ANN was constructed of three layers and used the learning rate of $1 \times 10^{-4}$.
By running WTMetaD of $\phi$ and $\psi$ variables with $F(\phi,\psi)$ as the potential energy, we could obtain $V_0^{\rm b}(\phi,\psi)$ as sum of gaussians.
Starting with this bias, we then carried out IMetaD for the alanine dipeptide with  $\phi$ and $\psi$ as CVs.
These runs were using a  $\gamma=5$, $w_0=0.3$~kcal~mol$^{-1}$,  $\delta s=0.25$~ rad, a deposition stride of 20~ps, and a time step of 2~fs.
In these runs all H-bonds in the system were constrained using the SHAKE protocol.

\subsection{Benzamidine Unbinding from Trypsin}

\begin{figure}
\begin{center}
\includegraphics[width=10.0cm]{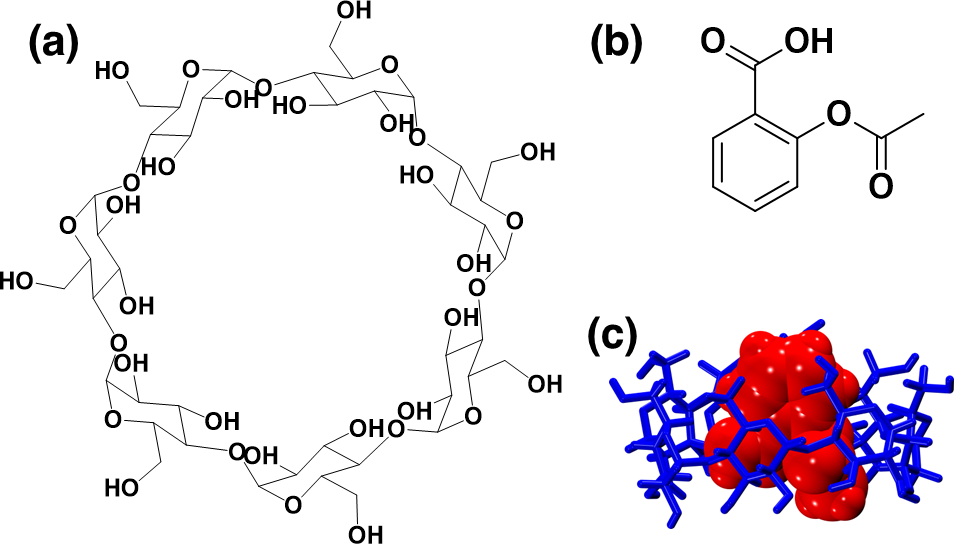}
\caption{ Structures of (a) $\beta$-CD, (b) aspirin and (c) $\beta$-CD aspirin complex}
\label{figure_combined}
\end{center}
\end{figure}

The initial structure was prepared from the bound trypsin-benzamidine complex crystal structure (PDB ID 1BTY);  \cite{katz1995episelection} see SI~Figure~S3.
The protein was modeled using the ff14SB AMBER force field \cite{maier2015ff14sb}, and benzamidine was parameterized with the GAFF force field.\cite{wang2004development}
The complex was solvated in a periodic box (65$\times$60$\times$70~{\AA}$^{3}$) containing 8100
 flexible TIP3P\cite{jorgensen1983comparison} water molecules with seven $\mathrm {Cl^{-}}$ anions added to neutralize the system.
 The box was sized so that, in the dissociated state, the ligand remained at least 15~{\AA} from the active site and its periodic images.
 After energy minimization, the system was equilibrated in the $NPT$ ensemble at 300~K and 1~bar, followed by further equilibration in the $NVT$ ensemble at 300~K. 
 Long-range electrostatic interactions were computed via the particle mesh Ewald (PME) method, and nonbonded interactions were evaluated with a 12~{\AA} cutoff.
MD simulations were performed using the %modified 
GROMACS-2019.4 code \cite{abraham2015gromacs} patched with %modified 
PLUMED-2.6.0. \cite{bonomi2009plumed,tribello2014plumed,plumed2019promoting}
For the equilibration of the system, we used a stochastic velocity rescaling thermostat \cite{bussi2009isothermal} with a relaxation time of 0.1~ps and Parrinello-Rahman barostat \cite{parrinello1981polymorphic} with isotropic pressure coupling at 1.0 bar with a time constant of 2.0~ps and compressibility of $4.5 \times 10^{-5}~ \text{bar}^{-1}$.

We performed bucket sampling type bias in our TASS simulations to study the dissociation of the ligand.
Three CVs were employed for the TASS simulations: (a) the distance (\textbf{Dis}) between C$_7$ (carbon having the diamine group) of benzamidine and C$_{\delta}$ of Asp189 in trypsin (See Figure S4); (b)  the coordination number (\textbf{Hbonds}) representing hydrogen bonds between selected trypsin active site residues (Asp189:OD1, Asp189:OD2, Val227:O, Val213:O, Tyr228:OH, Gly219:O, Ser190:O$_{\gamma}$) and selected atoms of benzamidine (N9, N10); (c) the water coordination (\textbf{LigSolv}) around selected set of atoms of benzamidine. 
See SI~Section~S2 for more details.
The \textbf{Dis} CV was used to apply bucket bias \cite{javed2024buckets} so that a controlled sampling of ligand unbinding could be achieved. 
The \textbf{Hbonds} CV enhanced the sampling of hydrogen bond formation and dissociation between protein side chains and the ligand, while the \textbf{LigSolv} CV accelerated the conformational sampling of solvent water molecules around the ligand. 
In the cases where the fluctuations of a CV were inherently small, a scaling was applied to facilitate oscillation along its corresponding auxiliary variable. 
The parameters $\mu$ and $\kappa$ for these CVs are provided in \tref{tab:table1}. 

\begin{table}
\caption{\label{tab:table1}{CVs used in the TASS simulation of benzamidine unbinding from trypsin and the corresponding auxiliary variable parameters $\mu$ and the $\kappa$ are listed.}}

\begin{ruledtabular}
\begin{tabular}{cccc}
CV & Scaling factor & $\mu$ (Da) & $\kappa$ (kcal mol$^{-1}$ {\AA}$^{-2}$) \\
\textbf{Dis} 	&	 20	& 0.05	& 478 \\
&  & $\mu$ (Da {\AA}$^2$) & $\kappa$ (kcal mol$^{-1}$) \\
\textbf{Hbonds}  	&	1 	&  4.73	& 4780 \\
\textbf{LigSolv}	&	 1	& 0.002	& 48\\
\end{tabular}
\end{ruledtabular}
\end{table}

The physical system was maintained at $T$=300~K  and the extended system at $\tilde T$=3000~K using a stochastic velocity rescaling thermostat \cite{bussi2009isothermal}, 
while the equations of motion were integrated with a 1~fs time step. 
A WTMetaD bias was applied along the \textbf{Dis} and \textbf{LigSolv} CVs. 
WTMetaD bias was using  $w_0=0.6$~kcal~mol$^{-1}$, a $\gamma=6$, and the bias deposition rate of 0.5~ps. 
The gaussian widths were 0.2~{\AA} for \textbf{Dis} and 0.5~(unitless) for \textbf{LigSolv} CV. 
The three dimensional free energy surface for the ligand dissociation was constructed from the TASS simulation  using the mean-force-based approach as in Ref.~\cite{pal2021mean}.

We used five bucket bias potentials along the {\bf Dis} CV to drive the dissociation process.
The ${\lambda}$ parameter, as in the bucket bias potential,  was determined to be 0.25~{\AA}, following the recipes in Ref.~\cite{javed2024buckets}.
Details of the bucket bias positions are summarized in \tref{tab:table2}.   

\begin{table}
\caption{\label{tab:table2}{Parameters related to the bucket bias potentials for the trypsin-benzamidine system.}}
\begin{ruledtabular}
\begin{tabular}{cccccc}
Windows & ${\xi}_h^{\rm L} ~({\text{\AA}}) $ & ${\eta}_h^{\rm L} ~({\text{\AA}})$ & ${\eta}_h^{\rm U} ~({\text{\AA}})$ & ${\xi}_h^{\rm U} ~({\text{\AA}})$  & $k_h$ (kcal mol$^{-1}$ {\AA}$^{-2}$)\\

1 	& 2.25 & 2.5 & 5.0 & 5.75 & 24\\
2 	& 4.75 & 5.0 & 7.5 & 7.75 & 24\\
3 & 7.25 & 7.5 & 10.0 & 10.25  & 24\\
4  	& 9.75 & 10.0 & 12.5 & 12.75 & 24\\
5	& 12.25 & 12.5 & 15.0 & 15.25 & 24\\
\end{tabular}
\end{ruledtabular}
\end{table}

We computed the projected free energy surface,  $F$(\textbf{Dis},\textbf{LigSolv}) on a $ 41 \times 41$ grid. 
We used this free energy surface for the IMetaD runs.
%we used the projected the free energy surface, $F(\textbf{Dis},\textbf{LigSolv})$.
%
Here, we used a 1~fs time step, a  $\gamma=10$, $w_0=0.63$ kcal mol$^{-1}$,  $\delta s$ values of 0.25~{\AA} and 0.25 (unitless) along 
$\textbf{Dis}$, $\textbf{LigSolv}$ CVs, 
respectively, and a deposition stride of 8~ps.
Rest of the procedures followed were identical to that discussed earlier.
\subsection{Aspirin Unbinding from $\beta$-Cyclodextrin}

The $\beta$-CD guest complex 
%is 
was
prepared by manually placing the guest inside the center of $\beta$-CD cavity; see \fref{figure_combined}. 
%Figure \ref{figure_combined}. 
%
The complex was then solvated in a cubic periodic box filled with TIP3P water molecules. \cite{jorgensen1983comparison}
For this system, 
$\beta$-CD was modeled using GLYCAM06, \cite{GLYCAM06} aspirin with Open Force Field Sage 2.0.0, \cite{openff1,boothroyd2023development,mobley2018escaping} and the simulation was carried out with a 1~fs time step. 
Long-range electrostatic interactions were calculated using the particle mesh Ewald (PME) method, and nonbonded interactions were evaluated with a 10~{\AA} cutoff. 
After an initial energy minimization, the solvated complex was equilibrated at 1~bar and 298~K using the Langevin middle thermostat \cite{zhang2019unified} and a Monte Carlo Barostat.\cite{aaqvist2004molecular} 
Subsequent equilibration was performed in the $NVT$  ensemble, and the simulations were executed using the OpenMM engine\cite{eastman2017openmm} and UFEDMM interface.\cite{bajpai2023interoperable,ufedmm}     

Umbrella bias based TASS simulations were performed to simulate host dissociation from the guest molecule.
Three CVs were chosen for these simulations: (1) the distance (\textbf{Dis}) between the center of mass of the $\beta$-CD and aspirin; (2) the number of contacts (\textbf{NContacts}) between the $\beta$-CD and the aspirin; (3) number of water molecules around the center of mass of aspirin (\textbf{LigSolv}).
See SI~Section~S3 for more details.
The {\bf Dis} CV was used to apply umbrella restraints for a directed sampling of the guest unbinding.
This is a critical CV along which the extent of unbinding can be monitored.
The {\bf NContacts} CV was chosen for enhancing the conformation of the guest molecule when it is in contact with the host. 
The {\bf LigSolv} is used to accelerate the conformational sampling of water molecules around the guest.
Ligand solvation has been reported to play an important role in accurately describing the unbinding event. \cite{limongelli2012sampling,casasnovas2017unbinding,perez2019local,rizzi2021role}

The scaling applied to the CVs, and the parameters $\mu_\alpha$ and $\kappa_\alpha$ parameters used in the TASS simulations are summarized in \tref{Asp_table}. 
The physical  system was maintained at 
 298~K while the extended system was kept at 1000~K.
 Temperatures were controlled using massive thermostatting with the Regulated- Nos\'e-Hoover-Langevin (R-NHL) thermostat \cite{abreu2021hamiltonian,leimkuhler2009gentle,leimkuhler2013robust}  with a time constant ($\tau$)=18~fs, friction coefficient $(\gamma)=10$~ps$^{-1}$, and regulation parameter $\text{n} =0.1$.  

\begin{table}
\caption{\label{Asp_table}{For the TASS simulation of $\beta$-CD-aspirin system, the CVs employed, their scaling factors, the $\mu_\alpha$ and the $\kappa_\alpha$ parameters are listed.}}
\begin{ruledtabular}
\begin{tabular}{cccc}
CV & Scaling factor & $\mu$ (Da) & $\kappa$ (kcal mol$^{-1}$ {\AA}$^{-2}$) \\

\textbf{Dis} 	&	 2000	& 0.002	& 0.05 \\
&  & $\mu_\alpha$ (Da {\AA}$^2$) & $\kappa_\alpha$ (kcal mol$^{-1}$) \\
\textbf{NContacts}  	&	1 	&  0.003	& 50.0 \\
\textbf{LigSolv}	&	 1000	& 0.005	& 0.5\\
\end{tabular}
\end{ruledtabular}
\end{table}

Umbrella restraints were placed along the \textbf{Dis} CV from 0.25~{\AA} to 15~{\AA} at intervals of 0.25~{\AA}.
An umbrella coupling constant $k_h
= 1 \times 10^{-3}$~kcal~mol$^{-1}$~{\AA}$^2$ was used for all the windows. 
TASS simulations were performed for 140~ns per window, and the free energy reconstruction was carried out using the mean-force-based approach.~\cite{pal2021mean}

For the IMetaD procedure, the two-dimensional projection of the free energy surface, $F(\textbf{Dis},\textbf{LigSolv})$ was used. 
We computed the free energy surface on a $ 60 \times 26$ grid.
The IMetaD runs were performed using a  $\gamma=6$, $w_0=0.5$~kcal~mol$^{-1}$, and $\delta s$ values of 0.25~{\AA} and 0.25 (unitless) along {\bf Dis} and {\bf LigSolv} CVs, respectively. 
A gaussian deposition stride of 8~ps was employed.
Rest of the procedures followed were identical to that discussed earlier.

\section{Results and Discussions}

\subsection{Alanine Dipeptide {\em In Vacuo}}

\begin{figure*}
\begin{center}
\includegraphics[width=15.0cm]{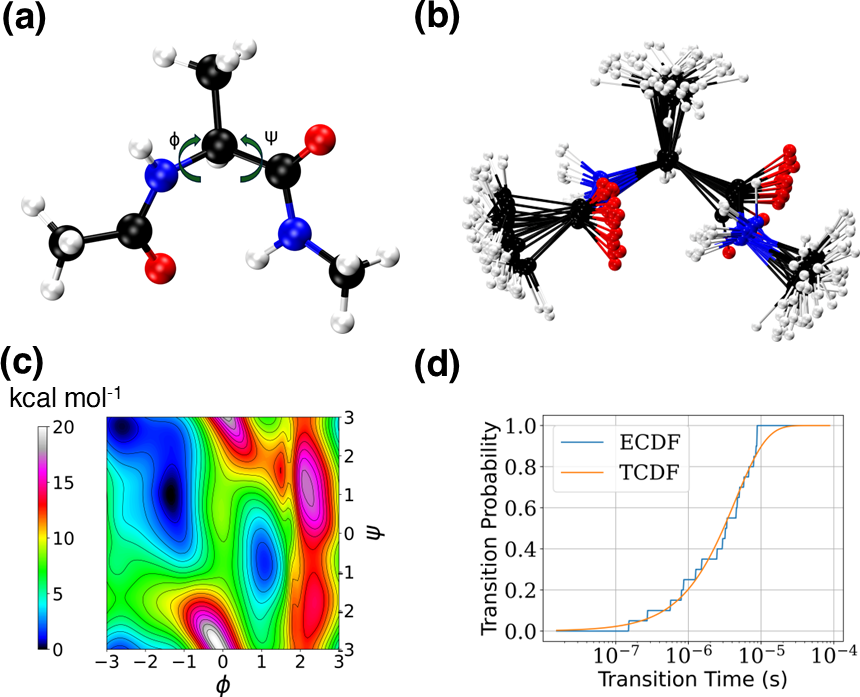}
\caption{Alanine dipeptide {\em in vacuo}: (a) C$_{7\mathrm{eq}}$ state, Atom colors: O (red), N (blue), C (black), H (white);
%$\phi$ (C-N-C$_{\alpha}$-C) and $\psi$ (N-C$_{\alpha}$-C-N) dihedral angles are shown. 
(b) Cluster of C$_{7\mathrm{ax}}$ state conformations from the  IMetaD simulations;
(c) Free energy surface $F(\phi,\psi)$ obtained from a trained ANN where the angles are in radians; (d) Transition probability.}
\label{ALA_combined}
\end{center}
\end{figure*}

To verify the approach proposed here, we took the
test case of alanine dipeptide  {\em in vacuo}. 
The two conformational states, C$_{7\mathrm{eq}}$ and  C$_{7\mathrm{ax}}$, are separated by a barrier of  $\sim$8~kcal~mol$^{-1}$. 
Two dimensional free energy surface, $F(\phi,\psi)$, was computed after 20~ns per window of TASS simulations, and an ANN was subsequently trained to represent this surface;  
See \fref{ALA_combined} and SI~Figure~S1.
The $L^2$ convergence of the free energy surface is provided in SI~Section~S1.

Next we obtained the bias potential $V_0^{\rm b}(\phi,\psi)$ which fills the basin of C$_{7\mathrm{eq}}$ conformational state nearly 90$\%$  using the protocols explained in Section~\ref{sec:imetad}. 
Subsequently, IMetaD was performed until the system reached the product basin, i.e., the C$_{7\mathrm{ax}}$ conformational state. 
The C$_{7\mathrm{ax}}$ conformational was identified in the IMetaD trajectories using the criteria $ 0.5 < \phi < 1.5 $ radians.
We carried out 20 independent simulations, to obtain $\tau^{\rm u}$.
We thus computed $\tau$ and $k$ for C$_{7\mathrm{eq}}$ to C$_{7\mathrm{ax}}$ transformation, 
and the computed value of $k=0.2 ~\mu \mathrm{s}^{-1}$
 is in good agreement with the previous studies.\cite{tiwary2013metadynamics,blumer2024short}
 The computed transition time is $\tau =4.4  ~ \mu \mathrm{s}$. 
 %with a standard deviation $\sigma =3.4 ~ \mu \mathrm{s}$. 
 The p-value from the KS test was found to be 0.84; See also Figure~\ref{ALA_combined} for the transition probability curves computed from the IMetaD simulations. 
\subsection{Benzamidine Unbinding from Trypsin}

\begin{figure*}
\begin{center}
\includegraphics[width=14.0cm]{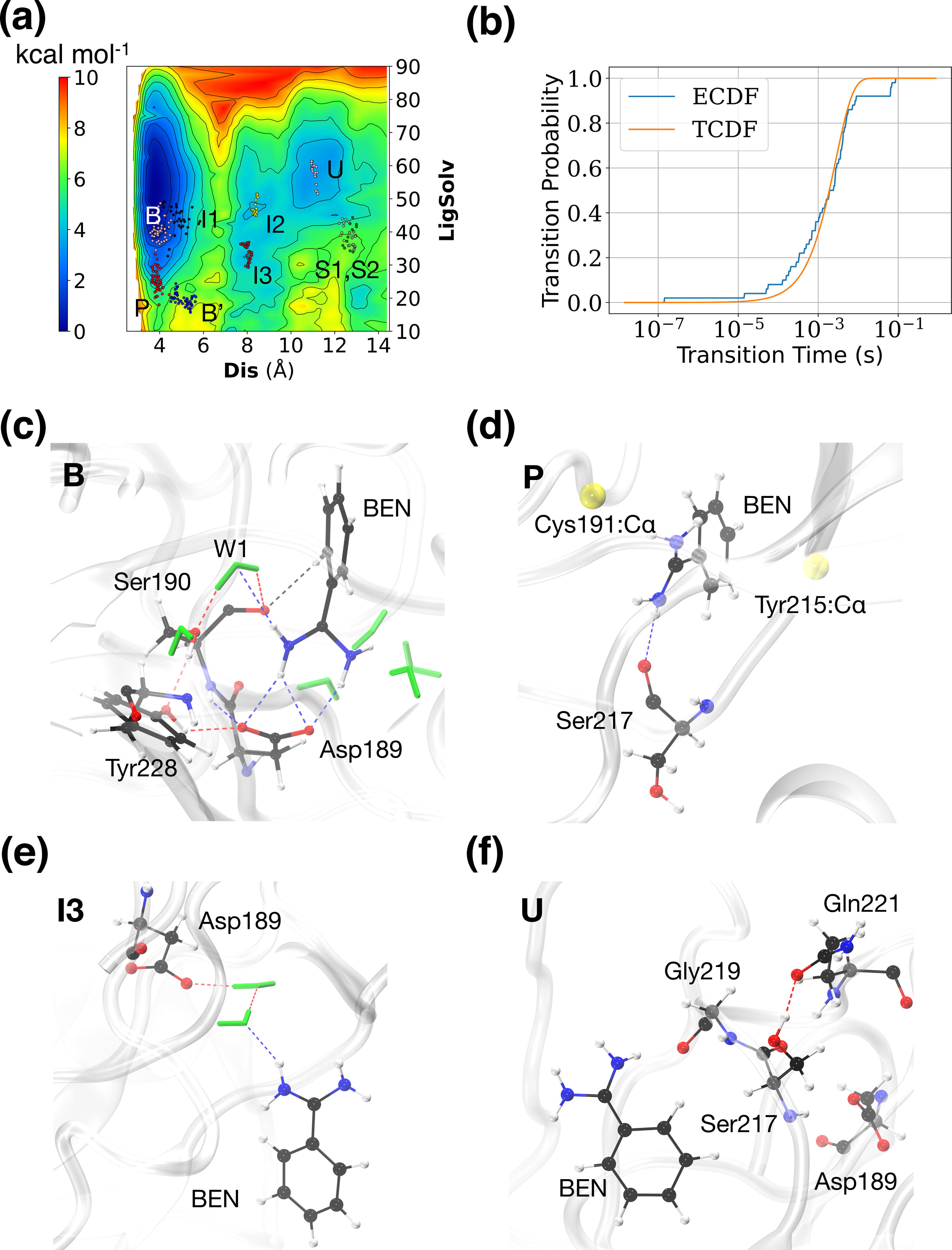}
\caption{Trypsin-benzamidine system: 
(a) Free energy surface $F$(\textbf{Dis},\textbf{LigSolv}) obtained from TASS simulations of benzamidine dissociation from trypsin is shown; (b) Transition probability computed from IMetaD simulations is provided; Various conformational states observed in the TASS simulations are also provided: (c) {\bf B}; (d) {\bf P}; (e) {\bf I3} (f) {\bf U}; See also SI~Figure~S8. Color code: C$_\alpha$ atoms of Cys191 and Tyr215 (yellow), water molecules (green).
}
\label{BEZ_combined}
\end{center}
\end{figure*}

We then investigated the 
%unbinding in the enzyme-inhibitor complex trypsin-benzamidine. 
unbinding of benzamidine from trypsin.
We computed the converged projected free energy surface $F({\bf Dis}, {\bf LigSolv})$ from the TASS simulations extending up to 500~ns per window; see \fref{BEZ_combined} and SI~Figures~S6 and S7.
We obtained $V_0^{\rm b}(\textbf{Dis}, \textbf{LigSolv})$ and IMetaD was performed using this bias, starting with the ligand-bound conformational state.
A total 50 independent IMetaD simulations were carried out till the ligand unbinding was observed.
Unbound state was characterized by {\bf Dis} CV having values greater than 10~{\AA}.
From each of the trajectories of these simulations, we obtained  $\tau^{\rm u}$ and thereby $\tau$ = 2.7 ms.
The KS test yielded a p-value of 0.61; See Figure~\ref{BEZ_combined} for the transition probability plot.
This corresponds to the rate constant $k \equiv k_{\mathrm{off}} = 365 ~ \mathrm{s}^{-1}$.
The computed $k_{\mathrm{off}}$ is in 
%excellent
agreement with the experimental value of $600 \pm 300 ~ \mathrm{s}^{-1}$ ~\cite{guillain1970use} and previous computations.\cite{{ansari2022water,tiwary2015kinetics,dickson2017multiple,buch2011complete,plattner2015protein,teo2016adaptive,doerr2014fly}}

In \fref{BEZ_combined}, we show the two-dimensional free energy surface $F(\textbf{Dis},\textbf{LigSolv})$ together with the locations of the conformational states; See also SI~Figure~S8. 
%(\fref{BEZ_intermediates}). 
%
Conformational states were defined by local minima on the free energy surface. Additional states described in the literature but lacking a corresponding minimum were identified by visual inspection of the trajectories. For each state, we ran short, unbiased simulations and projected the corresponding values of the CVs onto the free energy surface, sampled every 10 fs.
State {\bf B} is the global minimum of the free energy surface and matches with the X-ray binding pose (e.g. PDB 3atl) \cite{yamane2011crystal}.
In line with experiment \cite{schiebel2018intriguing} and prior simulation \cite{ansari2022water}, the W1 water molecule bridges the ligand and Ser190, and the Ser190 forms a hydrogen bond with Tyr228. 
Also, beneath Asp189, we observed on average five water molecules. 
In the states {\bf I2} and {\bf I3}, Asp189 and benzamidine are connected by a two-water bridge structure. 
State U denotes the unbound state, which is about at {\bf Dis}=11~{\AA} on the free energy surface. 
  
We also identified other conformational states.
In {\bf B$'$}, the diamino group of benzamidine forms a hydrogen bond with Ser190 and interacts with Asp189 via a water bridge; it also contacts Tyr228 and Val227, resembling the state {\bf B} in Ref.~\cite{tiwary2013metadynamics}. 
In state \textbf{I1}, a single water molecule connects Asp189 and benzamidine. Compared with state \textbf{B}, we observe increased hydration around Asp189, in line with Ref.~\cite{ansari2022water}.
In the state {\bf P}, the phenyl ring of benzamidine is sandwiched between the C$_\alpha$ atoms of Cys191 and Tyr215, engaging hydrophobic contacts, while the polar diamino group forms a hydrogen bond with Ser217, in agreement with Ref.~\cite{tiwary2013metadynamics}.

States {\bf S1} and {\bf S2} are observed when the ligand is unbound, in agreement with previous studies. \cite{buch2011complete,tiwary2013metadynamics}  
The {\bf S1} and {\bf S2} states differ in the existence of hydrogen bonding interactions between Gln221 and Ser217 or Gly219.

\subsection{Aspirin Unbinding from $\beta$-Cyclodextrin}
\begin{figure*}
\begin{center}
\includegraphics[width=15.0cm]{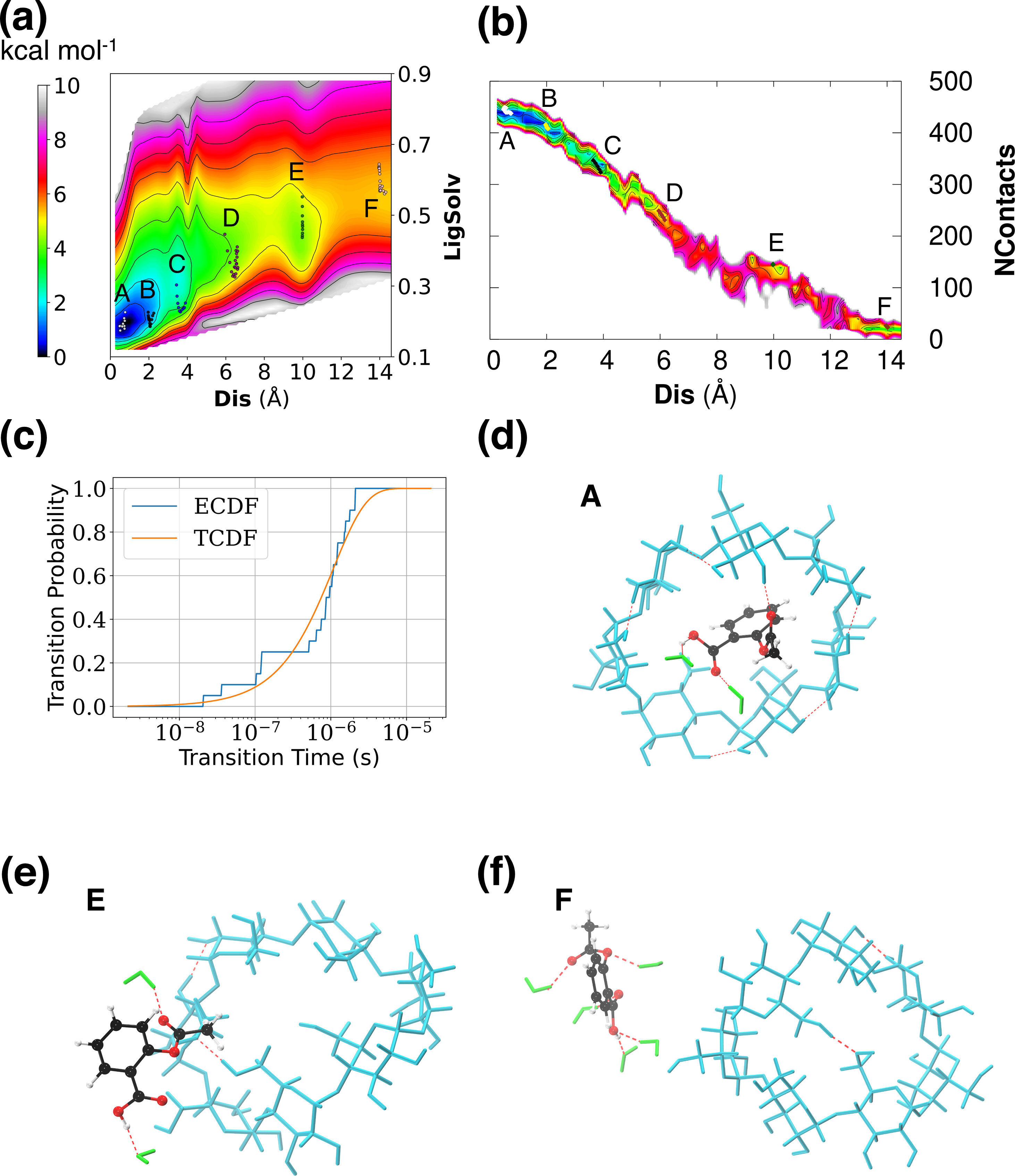}
\caption{(a) Free energy surfaces $F({\bf Dis},{\bf LigSolv})$
and (b) $F({\bf Dis},{\bf NContacts})$ from TASS simulation of dissociation of aspirin from $\beta$-CD; (c) Transition probability computed from the IMetaD simulations;  Various conformational states observed in the simulation are also reported: (d) {\bf A}; (e) {\bf E}; (f) {\bf F}; See SI~Figure~S14 for other conformational states identified. Color code: $\beta$-CD (cyan), water molecules (green), aspirin atoms are with C (black), O (red), and H (white).}
\label{HG_combined}
\end{center}
\end{figure*}

Finally, we studied the unbinding in the host-guest complex formed by $\beta$-CD and aspirin.
The three dimensional free energy surface 
$F({\bf Dis}, {\bf LigSolv}, \text{and} ~{\bf NContacts}$) was converged after 140~ns/window long TASS simulations (See SI~Figure~S12).
The projected free energy surface $F(\textbf{Dis}, \textbf{LigSolv})$ was then trained using an ANN; See Figure~\ref{HG_combined}.

Following the procedures laid out in this work, we computed  $V_0^{\rm b}(\textbf{Dis}, \textbf{LigSolv})$
that fills the free energy basin of the bound state by
90$\%$.
Using this $V_0^{\rm b}$ as the bias potential, we performed 20 independent IMetaD simulations, till the ligand is dissociated from the host.
The dissociation was characterized by {\bf Dis} CV with values more than 10~{\AA}.
From these simulations, we obtained  
$\tau = 1.0~\mu$s. 
The transition probability plot is in Figure \ref{HG_combined}; The KS test yielded a p-value of 0.65.
The above estimated value of $\tau$ can be translated into rate constant $k \equiv k_{\mathrm{off}}=1.0~\mu \mathrm{s}^{-1}$.
The computed rate constant is in good agreement with the experimental value of $1.3 \pm 0.03 ~ \mu \mathrm{s}^{-1}$ ~\cite{fukahori2006dynamic} and other computational studies. \cite{tang2018binding,votapka2022seekr2,miao2020ligand}

In \fref{HG_combined}, we show the two-dimensional free energy surface $F(\textbf{Dis},\textbf{LigSolv})$ and $F(\textbf{Dis},\textbf{NContacts})$ together with the locations of conformational states; see also SI~Figure~S14. 
We defined conformational states by the local minima of the free energy surface. Literature reported states lacking a minimum on this surface were assigned via visual trajectory inspection. For each state, short unbiased runs were performed, and the CV values were plotted on the free energy surface at 10 fs intervals.
State {\bf A} is the global minimum. 
In state {\bf B}, aspirin shifts towards one side of the cavity. 
In states {\bf C} and {\bf D}, it continues to slide towards the rim of the host. 
States {\bf A}-{\bf D} resemble to various states found in Ref.~\cite{ruzmetov2022binding}. 
While state {\bf E} has aspirin lies outside the cavity, but retains intermolecular contacts with the host, the {\bf F} denotes the unbound state, where aspirin move far from the rim region and diffuse in the solvent.

\section{Conclusion}

We introduced a computational approach to compute rate constants from TASS simulations using the ANN representation of free energy surfaces and the IMetaD protocol.
We demonstrated this for TASS simulations performed using both umbrella and bucket biases.
First, a low-dimensional projection of the TASS high-dimensional free energy surface is obtained, and this projection is then represented using an ANN. 
Next, a bias which fills up to 90\% of the barrier of the computed surface is constructed using a Langevin well-tempered metadynamics procedure.
Finally, IMetaD simulations were performed starting with this bias to obtain the kinetics.
This approach is simple to implement, and is shown to accurately predict barrier crossing rates for the conformational change of alanine dipeptide in \textit{vacuo}, benzamidine unbinding from trypsin, and  aspirin unbinding from $\beta$-CD. 
%
%In doing so, it complements the established capability of TASS to compute free energy surfaces while also extending its utility to kinetic predictions.
%
\begin{acknowledgments}
The manuscript is dedicated to Prof. N. Sathyamurthy on his 75th birthday. The authors acknowledge the National Supercomputing Mission (NSM) for providing the computational resources of PARAM Sanganak at IIT Kanpur. 
Authors thanks Mr. Shitanshu Bajpai and Dr. Shubhandra Tripathi for their help and valuable discussions.
Mr. Debjit Das acknowledges the SURGE fellowship by IIT Kanpur for his summer internship at IIT Kanpur during which a part of this research work was carried out.
\end{acknowledgments}

\section*{Data Availability Statement}

The data that support the findings of this study are available from the corresponding author upon reasonable request.

\FloatBarrier
%\nocite{*}
\bibliography{aipsamp}% Produces the bibliography via BibTeX.

%\clearpage
%\FloatBarrier

%\section*{Supporting Information: Kinetics of Barrier Crossing Events from Temperature Accelerated Sliced Sampling Simulations}

%\includepdf[]{SI.pdf}
%\appendix* 
\clearpage
\onecolumngrid
\section*{SUPPORTING INFORMATION}
\setcounter{section}{0}
\setcounter{figure}{0}
\setcounter{table}{0}
\setcounter{equation}{0}
\renewcommand{\thefigure}{S\arabic{figure}}
\renewcommand{\thesection}{S\arabic{section}}
\section{Alanine dipeptide in \textit{vacuo}}

\begin{figure}[htbp]
\begin{center}
\includegraphics[width=14.0cm]{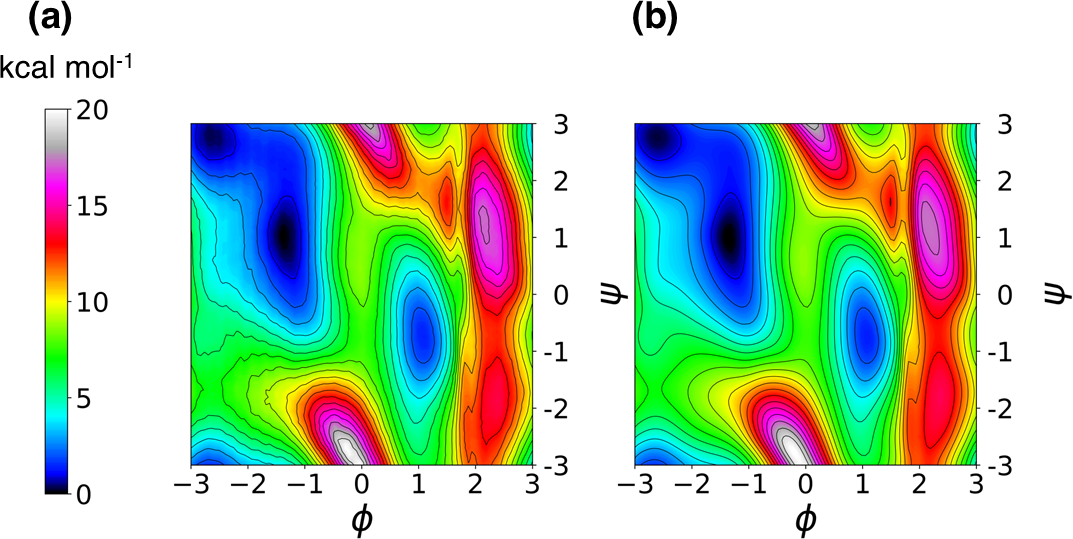}
\caption{Free energy surfaces, $F(\phi,\psi)$, for alanine dipeptide {\em in vacuo} (a) computed using TASS and (b) computed from the trained ANN.}
\label{ALA_NN}
\end{center}
\end{figure}

\begin{figure}[htbp]
\begin{center}
\includegraphics[width=12cm]{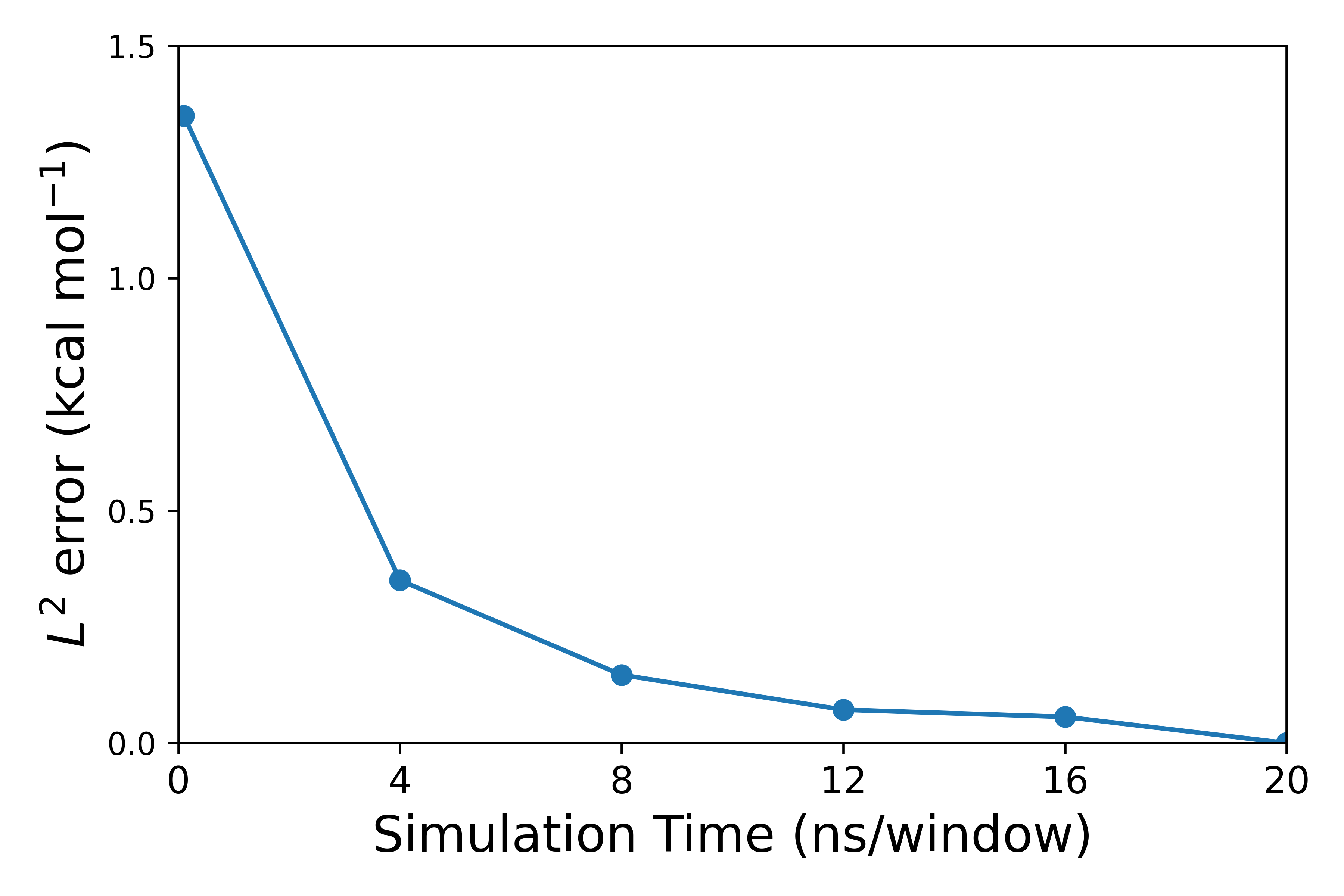}
\caption{Internal convergence of $F(\phi, \psi)$ for alanine dipeptide in \textit{vacuo} is monitored by computing the $L^2$ error by taking the 20~ns data as the reference.}
\label{L2_ALA}
\end{center}
\end{figure}

\section{Benzamidine Unbinding from Trypsin}

\subsection{Chemical Structure of Benzamidine}
\begin{figure}[htbp]
\begin{center}
\includegraphics[width=2.0cm]{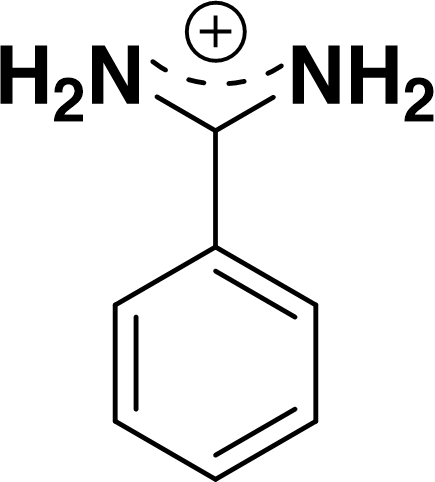}
\caption{Chemical structure of benzamidine}
\label{Structure_BEN}
\end{center}
\end{figure}
\subsection{Collective Variables}
\begin{enumerate}
\item \textbf{Dis}: Distance (\textbf{Dis}) between C$_7$ (carbon having the diamine group) of benzamidine and C$_{\delta}$ of Asp189 in trypsin (\fref{BEZ_Dis}).
%(Fig.~\ref{BEZ_Dis}). 
%
\begin{figure}[htbp]
\begin{center}
\includegraphics[width=8.0cm]{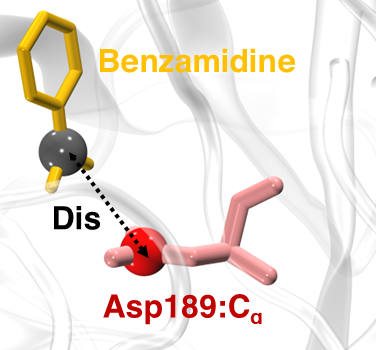}
\caption{ \textbf{Dis} CV is defined as the distance between C$_{\delta}$ (sphere, red) carbon of Asp189 and C$_7$ (carbon having the diammine group, shown as grey sphere) of benzamidine (stick representation, yellow).}
\label{BEZ_Dis}
\end{center}
\end{figure}
\item \textbf{Hbonds}: Coordination number of the trypsin active site residues (Asp189:OD1, Asp189:OD2, Val227:O, Val213:O, Tyr228:OH, Gly196:O, Ser190:O$_{\gamma}$) to a selected set of atoms of benzamidine (N9, N10) (\fref{BEZ_Hbonds_LigSolv}).
For atoms $i$ and $j$, the coordination number (CN) is defined as 
\begin{equation}
        \mathrm{CN}_{ij} = \frac{1-{(\frac{r_{{ij}}}{r_{{0}}}})^n}{1-{(\frac{r_{{ij}}}{r_{{0}}}})^m}
\end{equation}
where $r_{{ij}}$ is the distance between the atoms $i$ and $j$. 
Here, $r_\mathrm{0}$ is the distance cutoff.
We used $r_0=$3.5~{\AA}, 
$n = 6$, and $m = 12$. 
 \begin{figure}[htbp]
\begin{center}
\includegraphics[width=7.0cm]{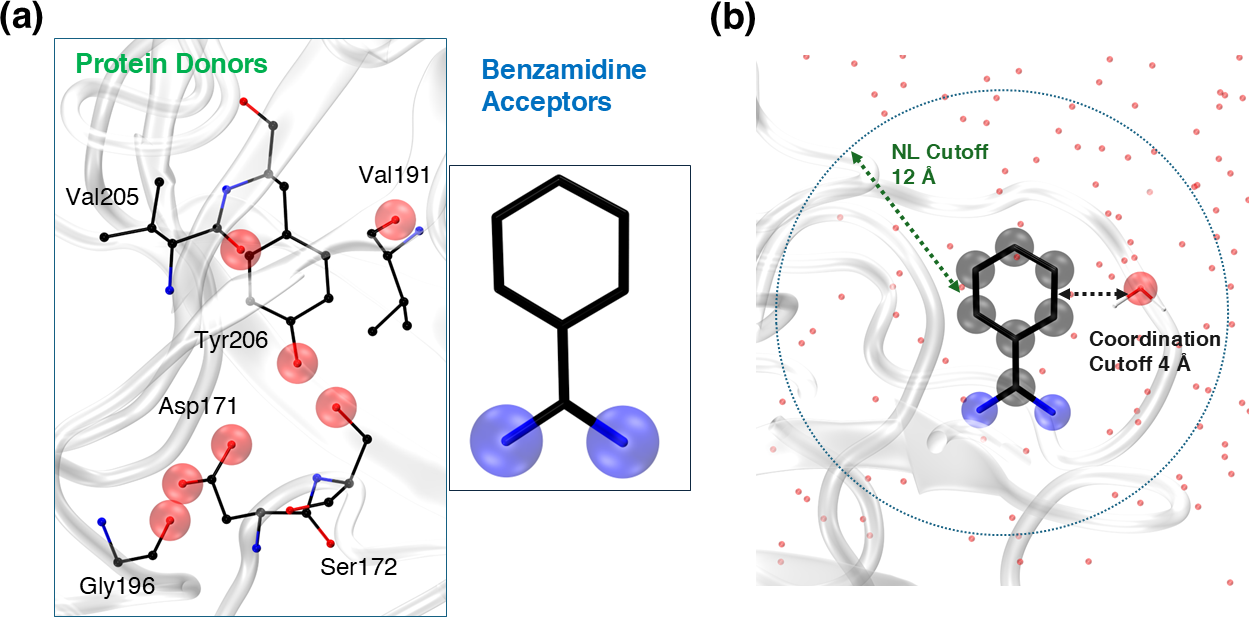}
\caption{(a) All the hydrogen-bond donor atoms (highlighted as spheres) of the trypsin active site residues (shown in ball and stick representations) and the corresponding hydrogen bond acceptor atoms (highlighted as spheres) of benzamidine (shown in ball and stick representations). Atom color codes: C (black), O (red), and N (blue). (b) Atoms of the benzamidine chosen for the \textbf{LigSolv} CV definition are shown (transparent spheres).}
\label{BEZ_Hbonds_LigSolv}
\end{center}
\end{figure}
\item \textbf{LigSolv}: Coordination number between benzamidine heavy atoms and water oxygen atoms; See \fref{BEZ_Hbonds_LigSolv}.
%Fig. \ref{BEZ_LigSolv}.
%
We chose $r_0=4$~\AA, and $n = 6$, and $m = 12$ in the definition of the CN function.
A neighbor list (NL) cutoff of 12 {\AA} was chosen, and the list updated at every 50 ps while computing the CN to speed-up the calculation.
We used $r_0=4$~{\AA}; 
A Neighbor list (NL) cutoff distance of 12~{\AA} taken. 
Atom color codes: C (black), O (red), and N (blue).  
\label{BEZ_LigSolv}
%\end{center}
%\end{figure}
\end{enumerate}

\subsection{Free Energy Surface}
\begin{figure}[htbp]
\begin{center}
\includegraphics[width=15.0cm]{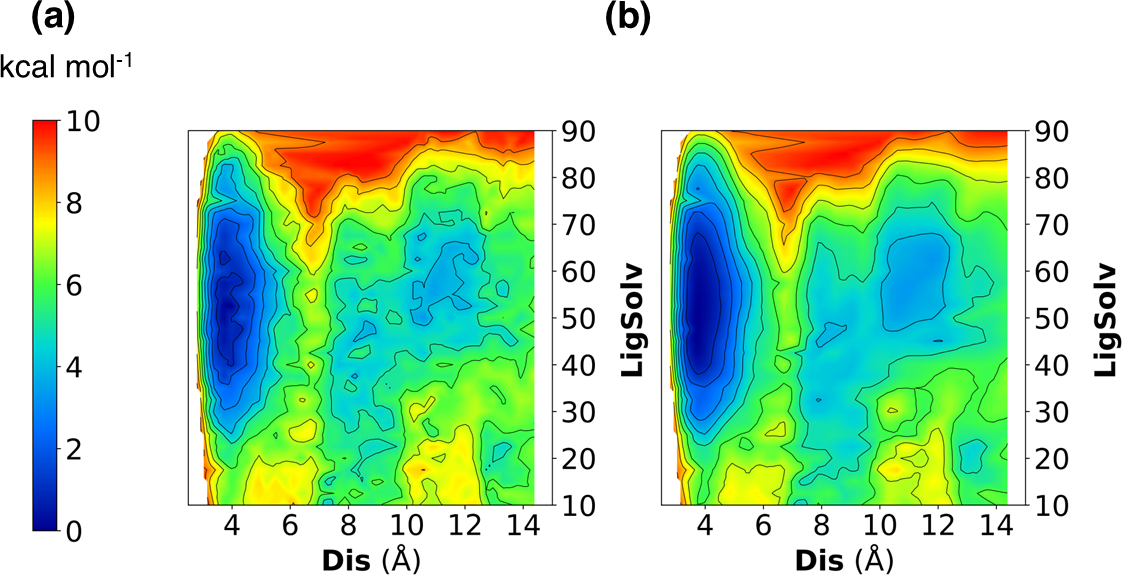}
\caption{Trypsin-benzamidine system: Free energy surface $F(
\textbf{Dis},\textbf{LigSolv})$
 as (a) computed using TASS and (b) from a trained ANN.}
\label{BEZ_NN}
\end{center}
\end{figure}
\begin{figure}[htbp]
\begin{center}
\includegraphics[width=10cm]{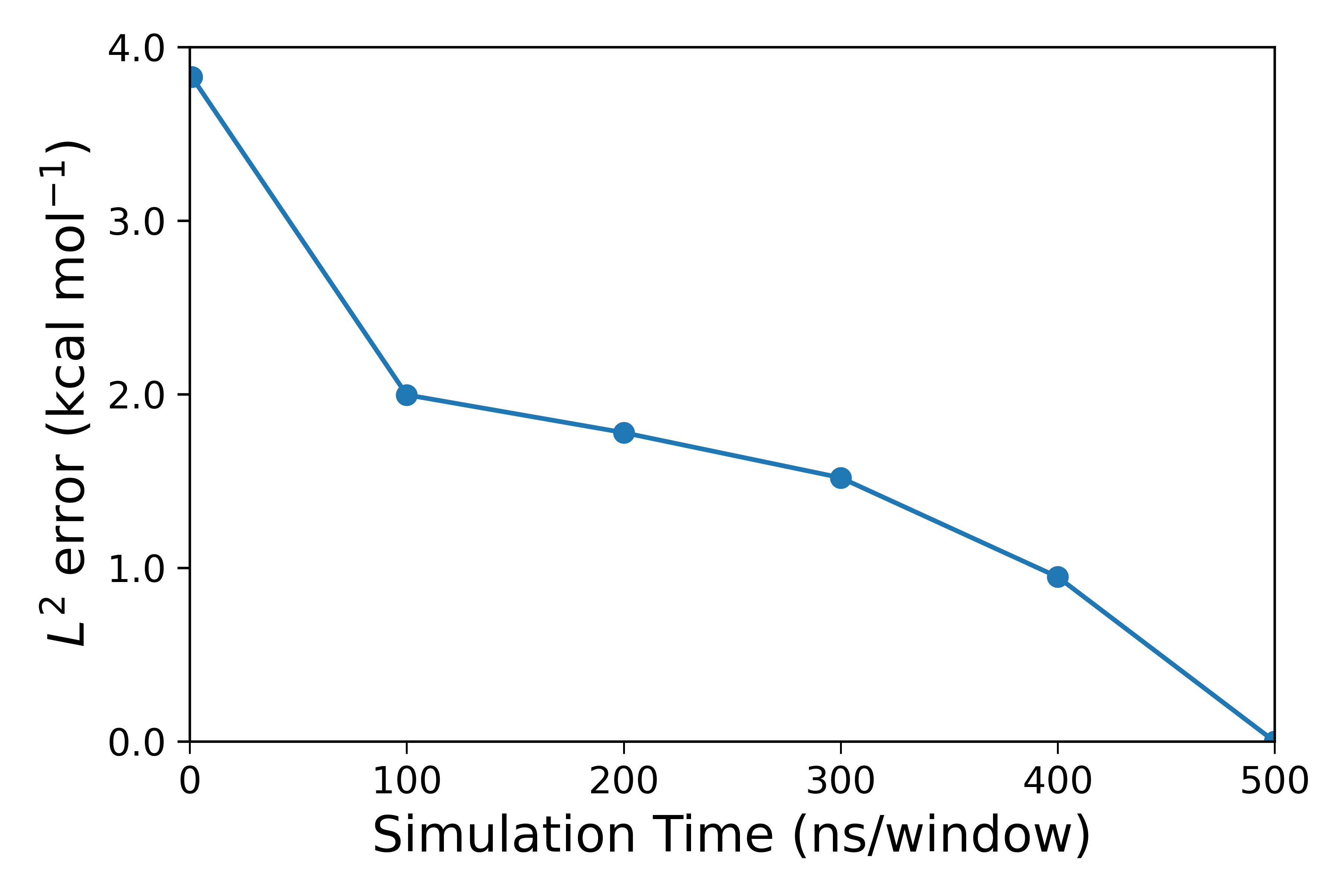}
\caption{Trypsin-benzamidine system: Internal convergence of 
$F$({\bf Dis}, {\bf Hbonds}, {\bf LigSolv}) monitored through 
the $L^2$ error. }
\label{L2u}
\end{center}
\end{figure}

\begin{figure*}
\begin{center}
\includegraphics[width=14.0cm]{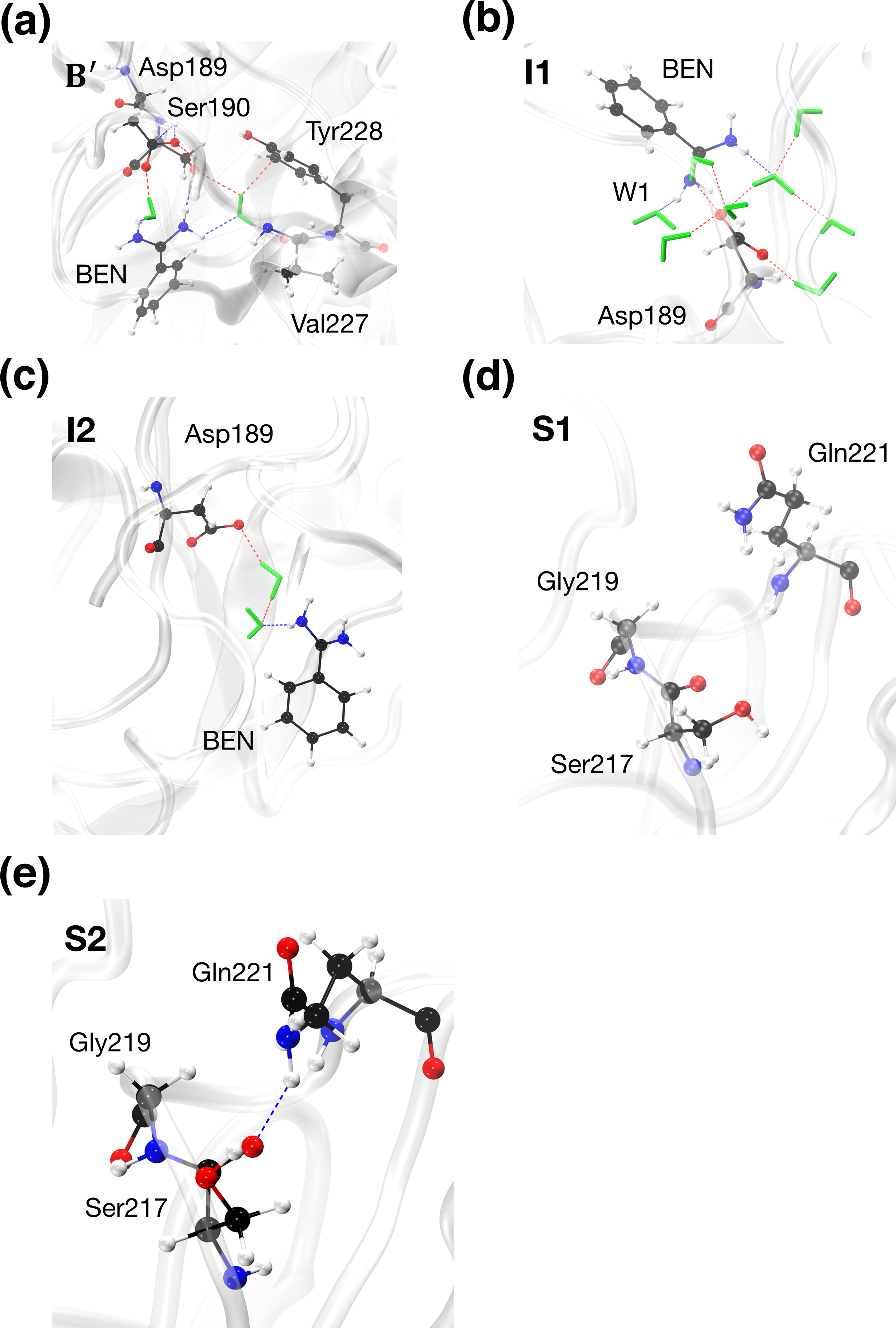}
\caption{
Various conformational states observed in the TASS simulation of benzamidine dissociation from trypsin are shown here: (a) {\bf B$^\prime$} ; (b) {\bf I1}; (c) {\bf I2}; (d) {\bf S1}; (e) {\bf S2}. Color code: water molecules (green).
}
\label{BEZ_intermediates}\end{center}
\end{figure*}
\clearpage

\section{Aspirin Unbinding from $\beta$-Cyclodextrin}
\subsection{Collective Variables}
\begin{enumerate}
%CV1
\item \textbf{Dis}: Distance between the center of mass of the host and the guest (\fref{HG_Dis}).
\begin{figure}[htbp]
\begin{center}
\includegraphics[width=8cm]{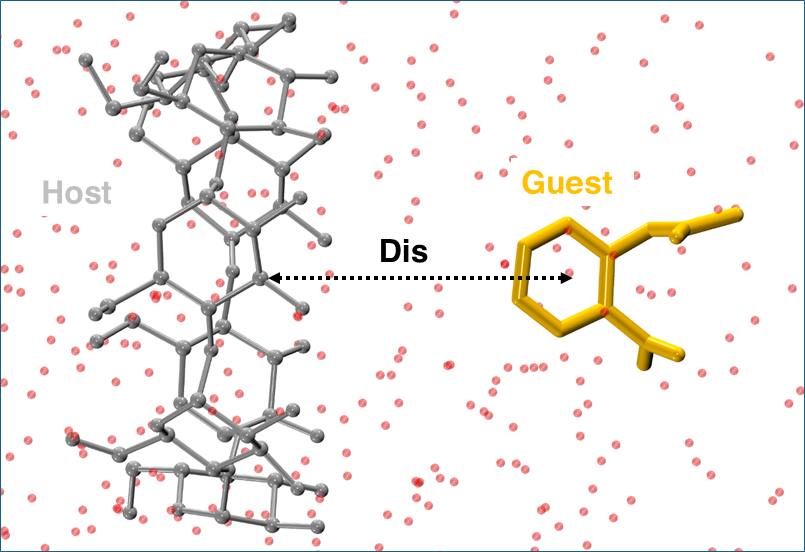}
\caption{\textbf{Dis} CV showing the distance between the center of mass of  $\beta$-CD (silver) and aspirin (yellow).}
\label{HG_Dis}
\end{center}
\end{figure}
% 
%CV2
\item \textbf{NContacts}: CN  between $\beta$-CD and aspirin ( \fref{HG_Contacts_figure}).
We used $r_0=6$~{\AA} and $n=6$ and $m=12$ in the definition of the CN function.
\begin{figure}[htbp]
\begin{center}
\includegraphics[width=10cm]{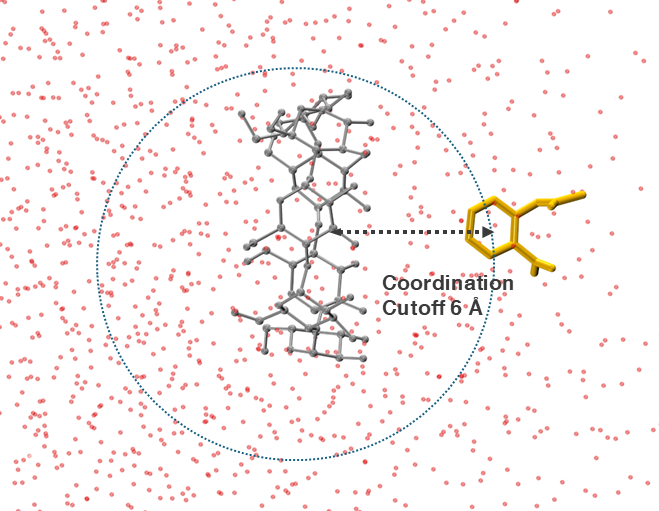}
\caption{Coordination of $\beta$-CD (silver) and aspirin (yellow) in the definition of the \textbf{NContacts} CV.}
\label{HG_Contacts_figure}
\end{center}
\end{figure}
%%CV3
\item \textbf{LigSolv}: Coordination number between center of mass of the guest and water oxygen atoms (\fref{HG_LigSolv_figure}).
Here we used $r_0=2.5$~{\AA}, $n=6$ and $m=12$ to define the CV. Atom color codes: C (black) and O (red).
\begin{figure}[htbp]
\begin{center}
\includegraphics[width=8cm]{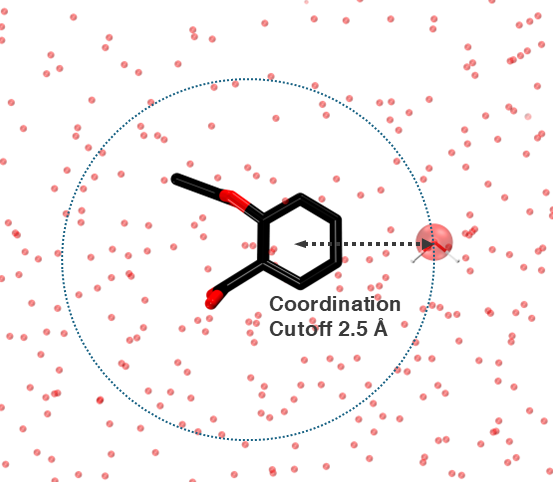}
\caption{Coordination between the center of mass of $\beta$-CD and  oxygen atoms of water molecules is defined as the \textbf{LigSolv} CV.}
\label{HG_LigSolv_figure}
\end{center}
\end{figure}
\end{enumerate}    
\clearpage
\subsection{Free Energy Surface}
\begin{figure}[htbp]
\begin{center}
\includegraphics[width=13.0cm]{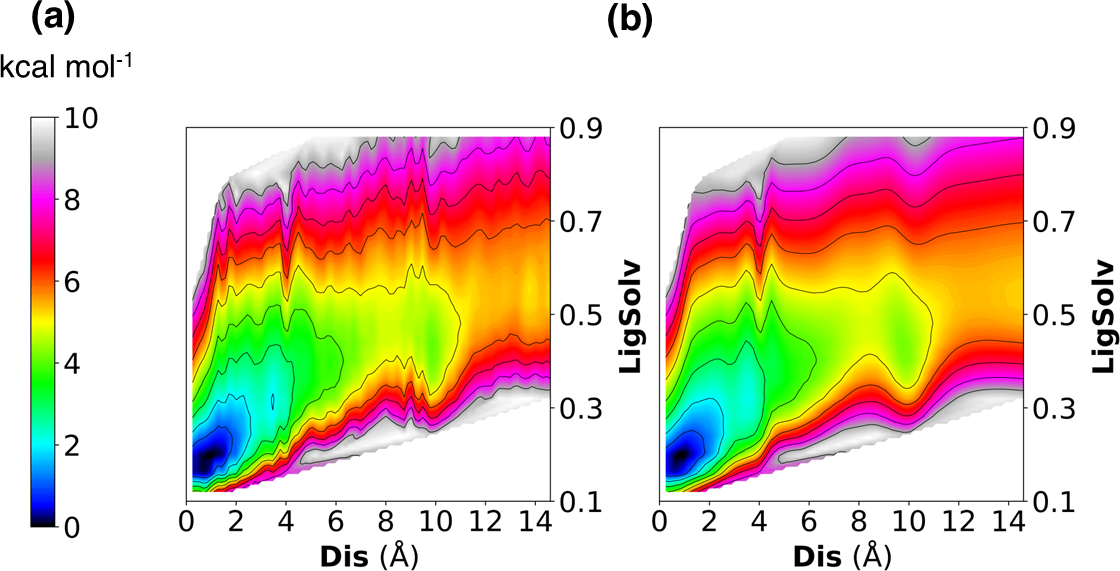}
\caption{$\beta$-CD-aspirin system: Free energy surface $F$(\textbf{Dis},\textbf{LigSolv}) (a) computed from TASS and (b) computed from a trained ANN.}
\label{HG_NN}
\end{center}
\end{figure}

\begin{figure}[htbp]
\begin{center}
\includegraphics[width=10cm]{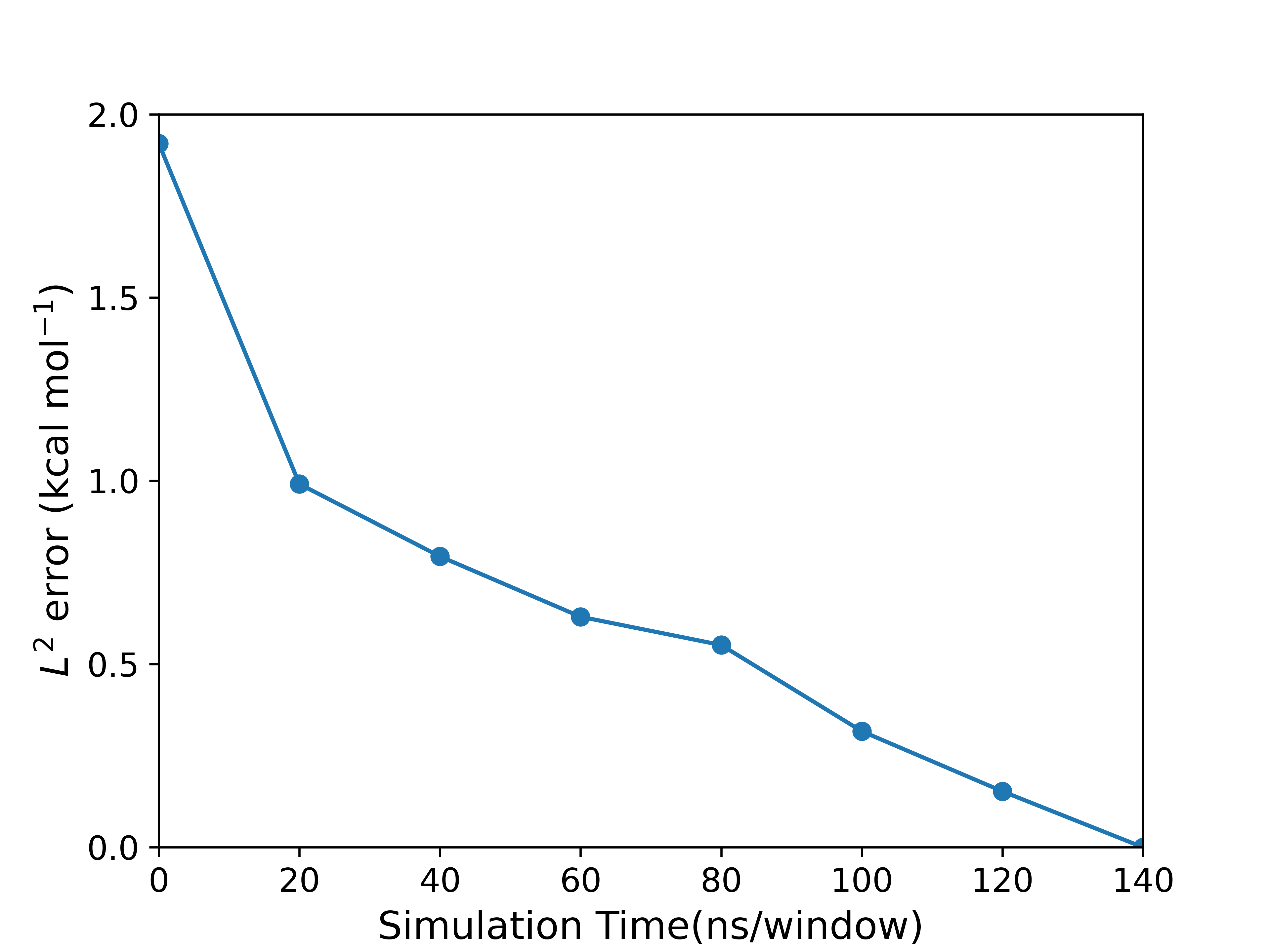}
\caption{$\beta$-CD-aspirin system:  Internal convergence of $F$(\textbf{Dis}, \textbf{NContacts}, \textbf{LigSolv}) monitored through $L^2$ error by taking the free energy surface at 140~ns as the reference.}
\label{L2u1}
\end{center}
\end{figure}

 \begin{figure*}
 \begin{center}
 \includegraphics[width=15.0cm]{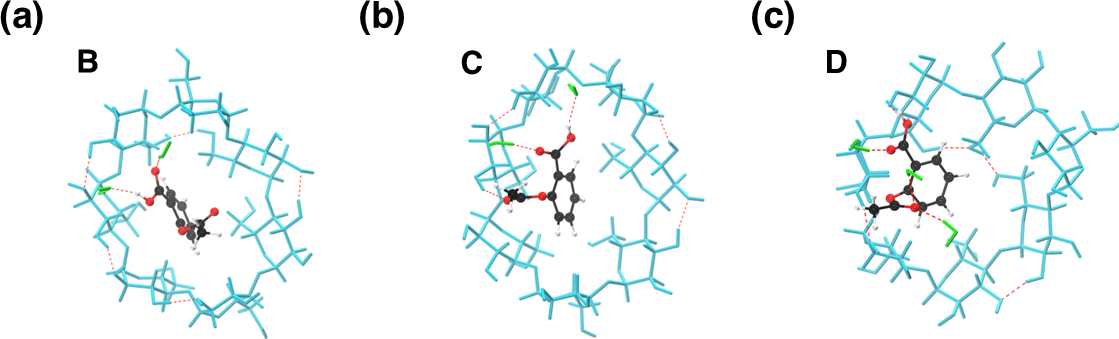}
 \caption{Various conformational states observed in the simulation of dissociation of aspirin from $\beta$-CD are shown here: (a) {\bf B}; (b) {\bf C}; (c) {\bf D}. Color code: $\beta$-CD (cyan), water molecules (green), aspirin atoms are with C (black), O (red), and H (white).
 }
 \label{HG_intermediates}
 \end{center}
 \end{figure*}

\end{document}